\documentclass[conference]{IEEEtran}
\IEEEoverridecommandlockouts
\usepackage{cite}
\usepackage{amsmath,amssymb,amsfonts}
\usepackage{algorithmic}
\usepackage{graphicx}
\usepackage{textcomp}
\usepackage{xcolor}
\usepackage{accessibility}
\usepackage{balance}
\usepackage[hidelinks]{hyperref}
\usepackage[caption=false,font=footnotesize]{subfig}
\def\BibTeX{{\rm B\kern-.05em{\sc i\kern-.025em b}\kern-.08em
    T\kern-.1667em\lower.7ex\hbox{E}\kern-.125emX}}
\begin{document}

\title{A Framework for Optimizing Human-Machine Interaction in Classification Systems\\
}

\author{\IEEEauthorblockN{1\textsuperscript{st} Goran Muric}
\IEEEauthorblockA{\textit{InferLink Corporation} \\
Los Angeles, CA, United States}
\and
\IEEEauthorblockN{2\textsuperscript{nd} Steven Minton}
\IEEEauthorblockA{\textit{InferLink Corporation} \\
Los Angeles, CA, United States}
}

\maketitle

\begin{abstract}
Automated decision systems increasingly rely on human oversight to ensure accuracy in uncertain cases. This paper presents a practical framework for optimizing such \emph{human-in-the-loop} classification systems using a \emph{double-threshold policy}. Conventional classifiers usually produce a confidence score and apply a single cutoff, but our approach uses two thresholds (a lower and an upper) to automatically accept or reject high-confidence cases while routing ambiguous instances to human reviewers. We formulate this problem as an optimization task that balances system accuracy against the cost of human review. Through analytical derivations and Monte Carlo simulations, we show how different confidence score distributions impact the efficiency of human intervention and reveal regions of diminishing returns, where additional review yields minimal benefit. The framework provides a general, reproducible method for improving reliability in any decision pipeline requiring selective human validation, including applications in entity resolution, fraud detection, medical triage, and content moderation.
\end{abstract}


\section{Introduction}
The motivation for this paper stems from a fundamental challenge in human-in-the-loop systems: when should a decision be deferred to a human expert, and how can we minimize human intervention while preserving high system performance (e.g., accuracy, precision, recall)?

We focus on a specific and practically relevant case -- classification systems that operate with a \emph{double–threshold policy}. 
Instead of a single cutoff, the automated classifier employs two thresholds, a lower ($\tau_l$) and an upper ($\tau_u$). Instances with scores below \(\tau_l\) are automatically assigned to one class, and those above \(\tau_u\) to the other, while scores falling between the two are considered uncertain and routed to human review. This scenario reflects a realistic operational scenario where human oversight over automated decision must exist.

Importantly, system performance is not determined by threshold choices alone (e.g., the F1 or accuracy achieved at a given cutoff). It also depends on the \textbf{distribution of predicted probabilities} for the incoming examples. In practice, these scores are often bimodal or highly skewed, which means that different operating regimes can expose very different mixes of “easy” and “ambiguous” cases. As a result, optimizing human involvement under such uncertainty becomes a non-trivial problem.

This challenge can be formulated as an optimization problem that calls for a rigorous theoretical framework. Consider an automated binary classifier that outputs a probability score between 0 and 1 for each input instance, indicating the likelihood that it belongs to the positive class. The goal is to decide, for each instance, whether to act automatically or defer the decision to a human. The resulting framework applies broadly to any binary decision task where probabilistic predictions are available and selective human oversight is required.

At the extremes of the score distribution, the automated decision is typically unambiguous: very low scores suggest the negative class while very high scores suggest the positive class. However, the difficulty lies in the intermediate region, where scores are neither clearly high nor clearly low. For these uncertain cases, the decision is escalated to human reviewers. When accuracy is critical, relying on a single decision threshold without human oversight may be insufficient, making a structured human-in-the-loop mechanism essential.  

The main objective, then, is to optimize the allocation of human effort: minimizing the number of instances requiring review while preserving (or ideally improving) overall system accuracy. Concretely, the framework takes as input a target objective and a review budget, and returns the threshold pair (\(\tau_l, \tau_u\)) on the Pareto frontier that optimizes the objective under that budget.

In this white paper, we introduce a formal mathematical framework to characterize this optimization problem. The framework can be applied broadly to classification systems where human validation is required for borderline cases. We further demonstrate its utility through Monte Carlo simulations, estimating optimal decision thresholds across a range of performance targets, including accuracy, precision, and recall. All accompanying code and simulations are fully reproducible and can be adapted by practitioners for arbitrary probability distributions.

\section{Related Work}
The idea of deferring uncertain decisions to a human or abstaining altogether has a long history in machine learning~\cite{hendrickx2024machine}. Recent research has formalized the problem of selectively deferring automated decisions to human experts under explicit cost models. A rigorous theoretical framework for classification with a reject option was developed by Bartlett and Wegkamp.~\cite{bartlett2008classification} They posed the problem as risk minimization with an explicit cost for rejection. Around the same time, El-Yaniv and Wiener~\cite{el2010foundations} formalized the notion of a selective classifier as a pair of functions: a predictor and a selection function that decides whether to output or abstain.

Rather than fixing a confidence threshold post hoc, one line of work optimizes the model to learn when to abstain as part of training. Cortes et al. introduced a formal predictor-rejector framework, in which a classifier and a rejector function are learned jointly to minimize a combined loss~\cite{cortes2023theory}. An alternative strategy is the classic score-based selective prediction, where one uses the classifier’s own confidence score (e.g. predicted probability) as a basis for deferral. This simpler approach has been widely used and was shown to be quite effective even for deep neural networks; Geifman and El-Yaniv later demonstrated that a neural network’s softmax probability can serve as a reliable uncertainty proxy.~\cite{geifman2017selective}

More recently, attention has shifted to explicitly human-in-the-loop classification and the question of optimal division of labor between humans and AI. Mozannar and Sontag introduced a consistent learning framework for learning to defer, where the classifier jointly learns both prediction and deferral strategies~\cite{mozannar2020consistent}. Follow-up research has extended this to scenarios with multiple experts or reviewers. For instance, Verma and Nalisnick~\cite{verma2022calibrated} study a calibrated one-vs-all scheme for deferring to the most suitable expert among many. Mozannar et al. in 2023~\cite{mozannar2023should} further tackled the optimization hardness of learning a joint human–AI system, proving that finding an optimal linear classifier and deferral rule is NP-hard in general

In summary, our approach builds on this rich body of work by focusing on a practical optimization of threshold policies for human oversight. While prior research has provided theoretical guarantees and learning algorithms for when to defer, there remains a need for operational guidance on setting the two critical thresholds in real-world deployments. Our contribution lies in quantitatively analyzing how different score distributions and performance targets influence the optimal double-threshold settings. Methodologically, our threshold sweep is related to work in human–computer interaction that characterizes performance as a continuous function of a threshold parameter rather than at a single cutoff, such as Vatavu's~\cite{vatavu2019} logistic-growth modeling of similarity tolerance and user consensus in gesture elicitation studies.

\section{Mathematical Formulation}\label{sec:math_formulation}
We formalize the human-in-the-loop classification process under a double-threshold policy, defining its data assumptions, decision rules, and expected outcomes.

\subsection*{Data}
We consider a generic binary decision problem with $N$ instances 
$\{x_i\}_{i=1}^N$. For each instance $x_i$, the classifier outputs a calibrated 
probability:
\[
p_i = \Pr(y_i=1 \mid x_i), \quad p_i \in [0,1],
\]
where $y_i \in \{0,1\}$ is the ground-truth label ($1=$ positive class, $0=$ negative class). Calibration implies that if $p_i = 0.95$, then in expectation 95\% of such instances truly belong to the positive class. For the sake of simplicity, we assume that the classifier itself does not introduce additional error beyond probabilistic uncertainty. That is, we treat $p_i$ as a perfectly calibrated estimate of the true conditional probability of a positive outcome.

\subsection*{Decision Rules}
The thresholds define how the system partitions decisions into automated and human-reviewed regions. 
Below, we formalize the expected outcomes for each action category. We define two thresholds:
\[
\tau_l \in [0,1], \quad \tau_u \in [0,1], \quad \tau_l < \tau_u.
\]

The decision rule for an instance $i$ is:
\[
a_i =
\begin{cases}
\text{Auto-Negative}, & p_i < \tau_l \\
\text{Review}, & \tau_l \leq p_i < \tau_u \\
\text{Auto-Positive}, & p_i \geq \tau_u.
\end{cases}
\]
We deliberately use the labels \emph{Auto-Negative} and \emph{Auto-Positive} rather than just ``Negative'' and ``Positive'' to distinguish system-generated decisions from outcomes that result from human review.

\subsection*{Outcomes and Expectations}
We now derive the expected outcomes associated with each decision category (\emph{Auto-Positive}, \emph{Review}, and \emph{Auto-Negative}) under the assumption of the calibrated probabilities. 

\begin{flushleft}
\textbf{Auto-Positive ($p_i \geq \tau_u$):} 
The system automatically assigns the positive class. The expected contribution of instance $i$ to true positives and false positives is:
\[
\begin{aligned}
\mathbb{E}[\mathrm{TP}_i \mid a_i = \text{Auto-Positive}] &= p_i, \\
\mathbb{E}[\mathrm{FP}_i \mid a_i = \text{Auto-Positive}] &= 1 - p_i, \\
\mathbb{E}[\mathrm{TN}_i \mid a_i = \text{Auto-Positive}] &= 0, \\
\mathbb{E}[\mathrm{FN}_i \mid a_i = \text{Auto-Positive}] &= 0.
\end{aligned}
\]

\textbf{Review ($\tau_l \leq p_i < \tau_u$):} 
The instance is sent to a human reviewer, whom we assume to be perfect. Each review incurs a unit cost. Thus,
\[
\begin{aligned}
\mathbb{E}[\mathrm{TP}_i \mid a_i = \text{Review}] &= p_i, \\
\mathbb{E}[\mathrm{TN}_i \mid a_i = \text{Review}] &= 1 - p_i, \\
\mathbb{E}[\mathrm{FP}_i \mid a_i = \text{Review}] &= 0, \\
\mathbb{E}[\mathrm{FN}_i \mid a_i = \text{Review}] &= 0.
\end{aligned}
\]

\textbf{Auto-Negative ($p_i < \tau_l$):} 
The system automatically assigns the negative class. Any true positive here becomes a false negative:
\[
\begin{aligned}
\mathbb{E}[\mathrm{TP}_i \mid a_i = \text{Auto-Negative}] &= 0, \\
\mathbb{E}[\mathrm{TN}_i \mid a_i = \text{Auto-Negative}] &= 1 - p_i, \\
\mathbb{E}[\mathrm{FN}_i \mid a_i = \text{Auto-Negative}] &= p_i, \\
\mathbb{E}[\mathrm{FP}_i \mid a_i = \text{Auto-Negative}] &= 0.
\end{aligned}
\]
\end{flushleft}




\subsection*{Global Quantities}

We define several aggregate measures that summarize the overall performance of the system under thresholds $(\tau_l, \tau_u)$.

\paragraph{Expected True Positives $C(\tau_l, \tau_u)$.} 
\begin{align}
C(\tau_l, \tau_u) &= \sum_{i=1}^N \big[ \mathbf{1}[p_i \geq \tau_u]\cdot p_i + \mathbf{1}
[\tau_l \leq p_i < \tau_u]\cdot p_i \big]
\end{align}
This is the expected number of correctly predicted positive instances under the thresholding strategy. It counts both automatically predicted positive pairs that are indeed true positives, and instances correctly identified by human reviewers in the uncertain region. In practice, $C$ captures the specific \textit{utility}\footnote{In settings where overall correctness is the appropriate goal, one can instead define utility as the total number of correct decisions,
$TP(\tau_l,\tau_u) + TN(\tau_l,\tau_u)$, or more generally as a cost-weighted combination of $TP$, $TN$, $FP$, and $FN$. Our framework and optimization procedure apply directly to these alternative objectives.} of the system—the higher its value, the more correct decisions are being made overall.

\paragraph{Expected False Positives $FP(\tau_l, \tau_u)$.} 
\begin{align}
FP(\tau_l, \tau_u) &= \sum_{i=1}^N \mathbf{1}[p_i \geq \tau_u]\cdot (1-p_i)
\end{align}
False positives arise only in the auto-accept region, where the system assigns positive classifications without human oversight. This quantity measures the expected number of incorrect positive predictions made automatically. In many domains (e.g., medical decisions, financial records, knowledge graphs), false positives may carry severe consequences. As such, $FP$ reflects the \textit{risk} of over-automation and must often be tightly controlled.

\paragraph{Expected False Negatives $FN(\tau_l, \tau_u)$.}
\begin{align}
    FN(\tau_l, \tau_u) &= \sum_{i=1}^N \mathbf{1}[p_i < \tau_l]\cdot p_i
\end{align}
False negatives occur when potentially positively classified instances are classified wrongly because their probability falls below the lower threshold $\tau_l$. This quantity measures the expected number of true positives that the system fails to recover. Depending on the application, the tolerance for $FN$ may vary: in recall-sensitive systems, minimizing $FN$ is critical.

\paragraph{Expected True Negatives $TN(\tau_l, \tau_u)$.}
\begin{align}
TN(\tau_l, \tau_u)
    &= \sum_{i=1}^N \mathbf{1}[p_i < \tau_u]\cdot (1 - p_i)
\end{align}
True negatives arise in two regions: (1) instances routed to human reviewers, where the reviewer perfectly identifies negative cases; and (2) instances automatically classified as negative when their score falls below the lower threshold $\tau_l$. No true negatives occur in the auto-positive region.

\paragraph{Expected Human Review Load $H(\tau_l, \tau_u)$.}  
\begin{align}
    H(\tau_l, \tau_u) &= \sum_{i=1}^N \mathbf{1}[\tau_l \leq p_i < \tau_u]
\end{align}
This term counts how many instances fall into the intermediate probability region and are thus sent to human reviewers. It reflects the \textit{workload burden} on human operators. Reducing $H$ is important when scaling the system. However, overly aggressive reductions in $H$ can lead to increased errors elsewhere (e.g., false positives or false negatives).

\subsection*{Optimization Objective (Fixed Review Budget)}
We focus on a single, practically motivated objective: \emph{maximize expected utility subject to a fixed review budget}. Let
\begin{align}
    \mathcal{F}(B) = \big\{\, (\tau_l, \tau_u) : 0 \le \tau_l < \tau_u \le 1,\; H(\tau_l,\tau_u) \le B \,\big\}
\end{align}

denote the feasible set under a review budget \(B\). More precisely, $\mathcal{F}(B)$ contains all pairs of thresholds $(\tau_l,\tau_u)$ for  which the expected number of items sent to human review does not exceed the allowed budget $B$. Our problem is
\[
\max_{(\tau_l,\tau_u)\in\mathcal{F}(B)} \; C(\tau_l,\tau_u),
\]
i.e., choose thresholds that maximize the expected number of correct decisions while ensuring the maximum load criterion.

This constrained formulation yields an \emph{accuracy–cost curve}: by varying \(B\), one traces a Pareto frontier that makes explicit the best achievable performance for any review capacity. In practice, this enables “budget-to-performance” planning and quantifies the marginal value of additional review resources.

Note that other objectives are possible, for example, a \emph{weighted cost minimization} that penalizes false positives, false negatives, and review effort through user-specified cost parameters. This formulation is particularly useful in domains where different error types have asymmetric consequences, such as medical diagnosis or fraud detection, as it allows organizations to encode domain-specific trade-offs directly into the optimization. We do not pursue this direction here, as it is beyond the scope of this paper, but the proposed framework can be readily adapted to support such extensions in future work.

\subsubsection*{Remark (Accuracy vs.\ $F1$ thresholds)}
When the objective is to maximize overall correctness under calibrated probabilities, the Bayes-optimal decision threshold is $0.5$. In this setting, the marginal value of review for an instance with score $p$ is $\Delta(p)=\min\{p,1-p\}$, which is maximized at $p=0.5$. Therefore,
under a fixed review budget, the optimal policy is to review the  instances closest to $0.5$, producing a symmetric double-threshold structure centered at $p=0.5$, regardless of the marginal distribution of scores.

The situation changes fundamentally when the performance objective is $F1$. The $F1$ score is a non-linear metric whose optimal decision rule does not coincide with the Bayes classifier for 0--1 loss. For calibrated classifiers, any threshold $t$ that maximizes $F1$ must satisfy $t = F1^/2 \le 0.5$, implying that the $F1$-optimal cutoff is generally \emph{below} $0.5$ and depends on class prevalence and the score distribution.~\cite{lipton2014optimal} As a result, the region in which human review  yields the largest marginal gain for $F1$ is no longer centered at $0.5$ and may be asymmetric around $t$. 

In summary, with accuracy-based objectives the optimal threshold is usually centered around $0.5$. However, $F1$ optimization induces a different and potentially asymmetric review region. This distinction is important when interpreting the threshold landscapes and Pareto frontiers in our simulation results, which consider both accuracy-like and $F1$-based objectives.

\section{Simulation Studies}
We conducted \textbf{Monte Carlo simulations} to estimate the optimal decision thresholds under diverse performance objectives and cost configurations. 
For each configuration, we computed empirical estimates of the expected cost 
\( C(\tau_l, \tau_u) \), false positives 
\( FP(\tau_l, \tau_u) \), and human review workload 
\( H(\tau_l, \tau_u) \). 
We then derived the corresponding \textit{Pareto-optimal frontiers} that capture the trade-offs between these quantities across the parameter space of \( (\tau_l, \tau_u) \).

\subsection*{Probability Distributions}

To conduct the simulations, we must first define the probability distribution representing the output of the system's internal classifier. 

In many practical applications, especially those with well-calibrated models and clear separation between classes, binary classifiers tend to produce probabilities concentrated near 0 or 1, with relatively few values in the intermediate range. 
In our own entity resolution systems, for example, we observe that most candidate pairs are either very likely matches or very likely non-matches, leading to a strongly bimodal distribution of scores.\footnote{To verify this behavior, we extracted all matching probabilities from our internal entity resolution model and indeed observed a strongly bimodal distribution.}

Based on this observation, we model the classifier's output probabilities using a bimodal mixture of two Beta distributions -- one peaking near 0 and the other near 1. 

In real-world settings, however, classifier outputs are often unbalanced, favoring either high or low probability regions. 
To capture this variation, we define two additional skewed distributions, resulting in a total of three distributions used in our simulations:
\begin{enumerate}
    \item \textbf{Beta mixture} -- a balanced bimodal distribution composed of two Beta components, one peaking near 0 and the other near 1, representing a classifier that produces both confident positives and negatives;
    \item \textbf{Beta Right Skewed} -- a right-skewed distribution with most probability mass near 1, corresponding to a system that generates predominantly high-confidence (positive) predictions; 
    \item \textbf{Beta Left Skewed} -- a left-skewed distribution with most probability mass near 0, corresponding to a system that tends toward low-confidence or negative predictions.
\end{enumerate}

\begin{figure}
    \centering
    \includegraphics[width=0.95\linewidth]{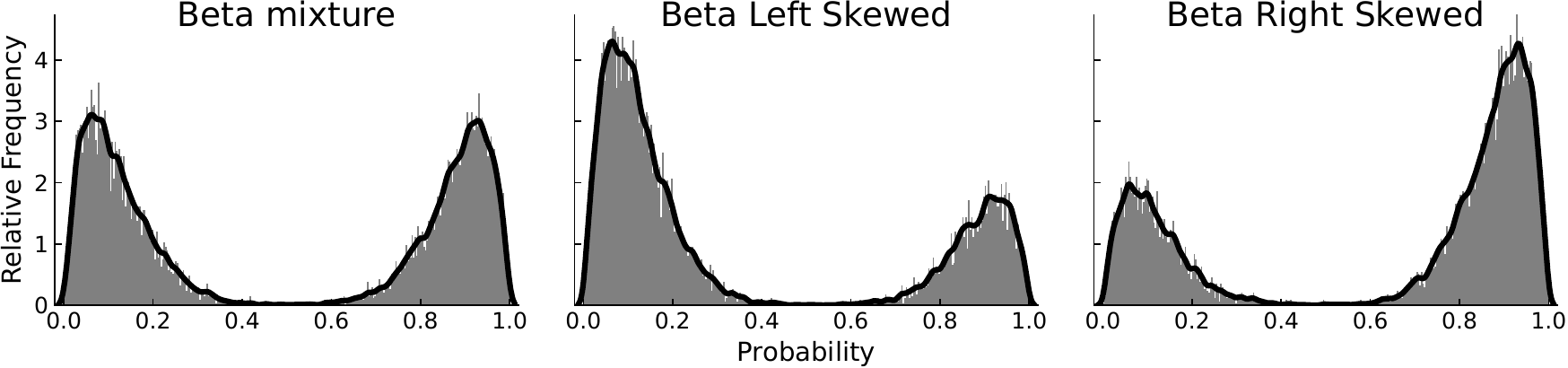}
    \caption{Density plots of the simulated probability distributions used in the Monte Carlo experiments. The \textit{Beta mixture} shows balanced bimodality, while the \textit{Beta Right Skewed} and \textit{Beta Left Skewed} distributions are skewed toward high and low probabilities, respectively.}
    \label{fig:probability_distributions}
\end{figure}

These distributions collectively allow us to explore a range of classifier behaviors -- from balanced uncertainty to strong bias toward one class. 
Figure~\ref{fig:probability_distributions} shows the estimated probability density functions for each case, highlighting how the concentration of predicted probabilities affects the expected number of ambiguous cases requiring human review.

\subsection*{Simulation Setup}

To empirically explore the trade-offs in human-in-the-loop classification, we designed a Monte Carlo simulation framework that systematically varies the decision thresholds and underlying probability distributions. 
Each simulation scenario consists of \( N = 10{,}000 \) entities, each assigned a probability value sampled from one of the three distributions described in the previous section.  

We define two decision thresholds: the lower threshold \( \tau_l \) and the upper threshold \( \tau_u \). 
Probabilities below \( \tau_l \) are automatically classified as negative, while those above \( \tau_u \) are classified as positive. 
Probabilities falling between these two thresholds are deferred to human reviewers.  

To explore the parameter space, we sample: 30 values of \( \tau_l \) uniformly spaced between 0.01 and 0.50; and 30 values of \( \tau_u \) uniformly spaced between 0.50 and 0.99.

For each combination of thresholds \( (\tau_l, \tau_u) \), we repeat the simulation 100 times to obtain robust empirical estimates of the expected values. 

\subsubsection*{Choice of Beta parameters.}
The Beta distributions used in our simulations are selected to model realistic classifier confidence patterns. We use $\mathrm{Beta}(15,2)$ to represent a high-confidence positive mode: this distribution is strongly concentrated near~1. For the high-confidence negative mode, we use $\mathrm{Beta}(2,15)$, with mass concentrated near~0. Using mixture weights of $(0.5,0.5)$ yields a balanced bimodal shape, while asymmetric weights such as $(0.7,0.3)$ or $(0.3,0.7)$ shift the distribution toward predominantly high or low confidence regions. 

\subsubsection*{Estimation of $F1$.}
To approximate the expected $F1$ under each operating point $(\tau_l,\tau_u)$, we use a Monte Carlo procedure. For each pair of thresholds and each probability distribution, we run $R=100$ simulation runs. In run $r$, we compute the confusion-matrix counts $\mathrm{TP}^{(r)}, \mathrm{FP}^{(r)}, \mathrm{FN}^{(r)}$ over $N = 10{,}000$ instances and evaluate
\[
F1^{(r)} =
\frac{2\,\mathrm{TP}^{(r)}}{2\,\mathrm{TP}^{(r)} + \mathrm{FP}^{(r)} + \mathrm{FN}^{(r)}}.
\]
We then report the Monte Carlo estimate
\[
\widehat{\mathbb{E}}[F1] = \frac{1}{R} \sum_{r=1}^R F_1^{(r)},
\]
which approximates the \emph{expected} $F1$ score, i.e., the expectation of the ratio.

\subsection*{Simulation Results}
We now examine the outcomes of the Monte Carlo simulations under the human-in-the-loop classification scenario. 

\subsubsection*{Threshold Sensitivity}
\begin{figure}[htbp]
    \centering
    \includegraphics[width=0.95\linewidth]{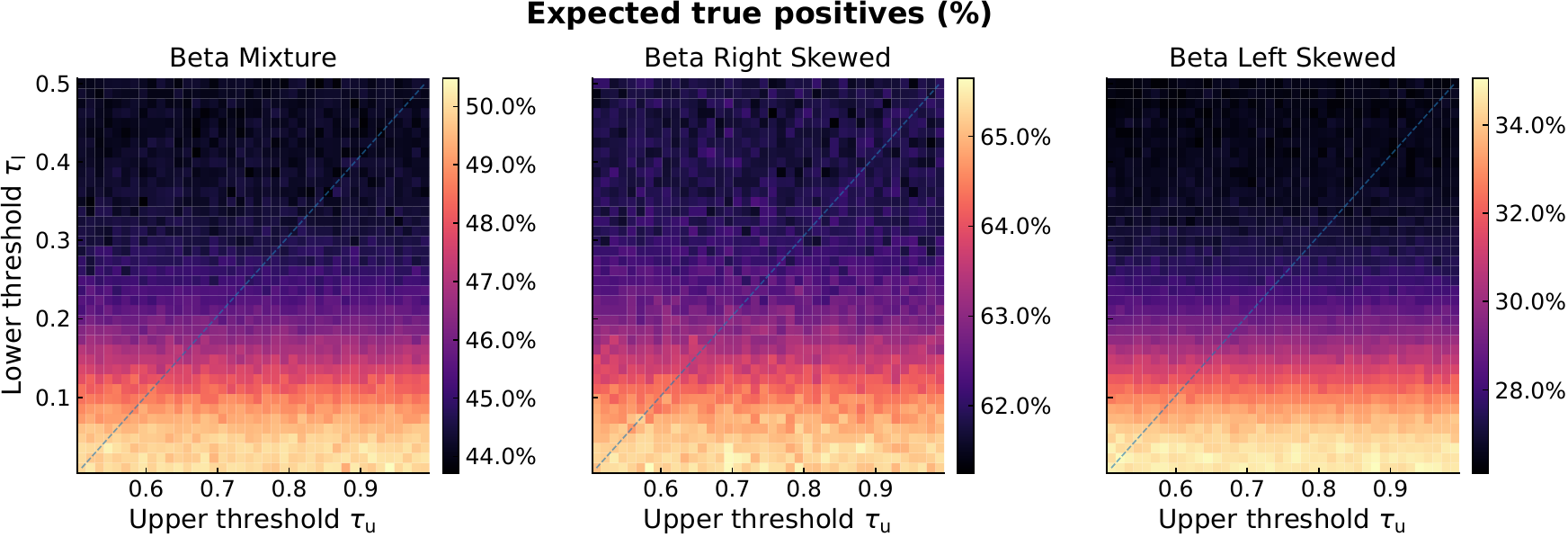}
    \caption{\textbf{Expected true positives (\%)} as a function of the lower ($\tau_l$) and upper ($\tau_{u}$) thresholds across three simulated score distributions: \emph{Beta Mixture}, \emph{Beta Right Skewed}, and \emph{Beta Left Skewed}. Each panel visualizes the expected percentage of correctly classified pairs for a unique ($\tau_{l}$, $\tau_{u}$) operating point. Color scales are panel-specific.}
    \label{fig:total_triptych}
\end{figure}

Figure~\ref{fig:total_triptych} presents the heatmaps of the expected fraction of \textit{true positives}, denoted as the percentage of correctly classified pairs across the simulation space of thresholds \( (\tau_l, \tau_u) \). 
Each point on the heatmap corresponds to a unique combination of lower and upper thresholds, indicating the boundary conditions under which automated or human-reviewed decisions are made.

Lowering $\tau_l$ assigns more borderline low-score pairs to human review rather than auto-rejecting them, thereby increasing the total number of correct outcomes. In contrast, the metric is comparatively insensitive to $\tau_u$ over the explored range; with bimodal score distributions, most true positives already concentrate near 1, so shifting the upper cutoff has limited effect on this measure.

The surface has the same shape across all three simulated score distributions. Only the absolute levels differ (low in \textit{Beta Right Skewed}, mid in \textit{Beta Mixture}, high in \textit{Beta Left Skewed}), which should not affect the threshold policy.

\begin{figure}[htbp]
    \centering
    \includegraphics[width=0.95\linewidth]{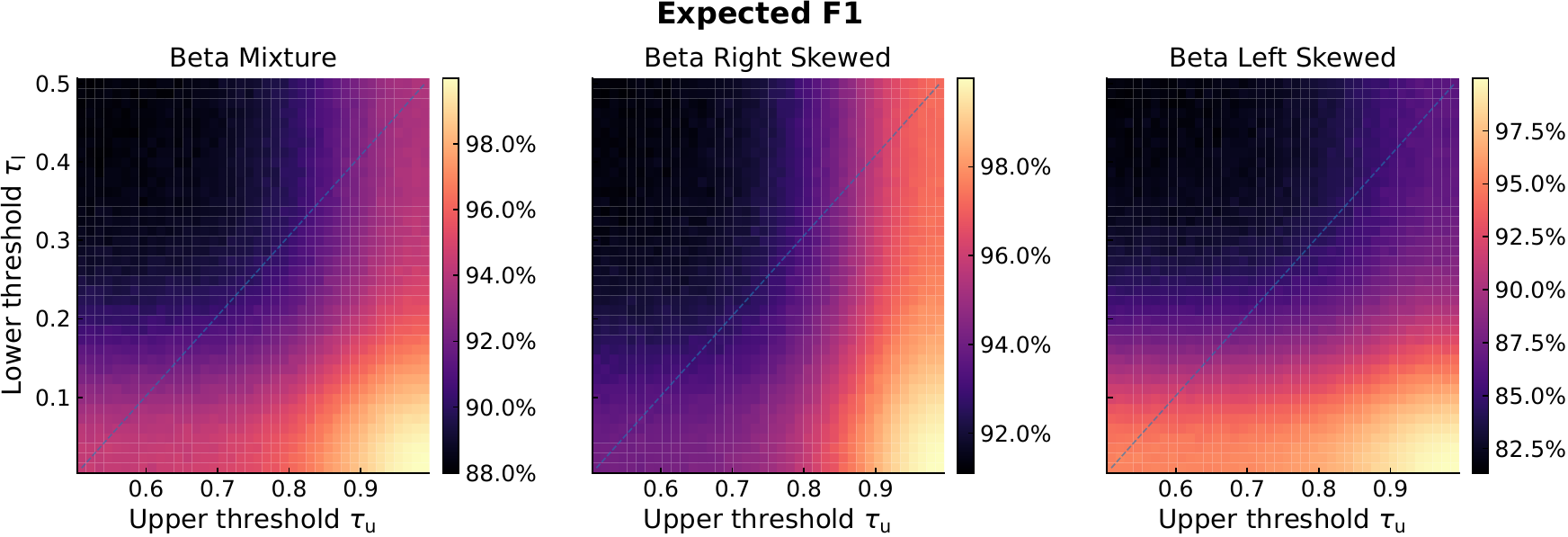}
    \caption{\textbf{Expected F1} as a function of the lower ($\tau_l$) and upper ($\tau_u$) thresholds under three simulated score distributions. Each panel visualizes the expected F1 score for a unique ($\tau_{l}$, $\tau_{u}$) operating point. Color scales are panel-specific.}   
    \label{fig:f1_triptych}
\end{figure}

Similarly, Figure~\ref{fig:f1_triptych} reports the expected F1 scores across all ($\tau_l$,$\tau_u$) combinations for the three simulated score distributions.

Under the \textit{Beta–mixture setting}, the expected F1 increases monotonically with the upper cutoff \(\tau_u\) and with lowering the lower cutoff \(\tau_l\). The best operating region is the \emph{lower–right} corner (high \(\tau_u\), low \(\tau_l\)), which accepts only the most confident positives while routing borderline low scores to human review. In practice, pushing \(\tau_u\) high will mostly maximize F1. 

Relative to this baseline, the \textit{Beta Right Skewed} regime
shows a stronger dependence on \(\tau_l\). Conversely, in the \textit{Beta Left Skewed} regime the sensitivity flips: F1 is dominated by \(\tau_u\), whereas varying \(\tau_l\) has limited
effect. Operationally, maximizing F1 would highly depend on the underlying probability distribution.

Additional results on expected recall, precision, false negatives, false positives, and human review workload (Figures~\ref{fig:RC_triptych} --\ref{fig:human_load_triptych}) are provided in Appendix.

\subsubsection*{Optimal Operating Boundary}

\begin{figure}[h]
    \centering
    \includegraphics[width=0.80\linewidth]{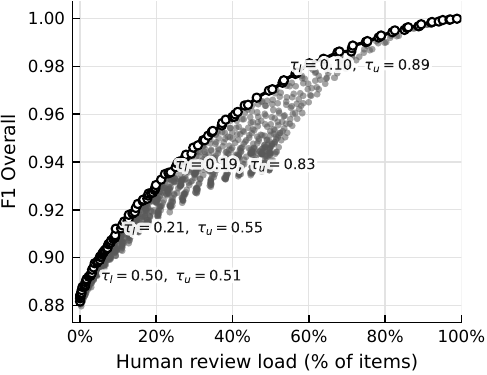}
    \caption{\textbf{Pareto frontier of F1 vs. human review load.}
    Each gray dot is one operating point defined by a pair of thresholds ($\tau_l$,$\tau_u$). The outlined curve traces the \emph{Pareto frontier}: the optimal settings that achieve the highest F1 for a given review budget. Example operating points are annotated with their ($\tau_l$,$\tau_u$) values. The analysis is conducted under the Beta–mixture score regime.}
    \label{fig:workload_vs_f1}
\end{figure}

We also analyze the \emph{optimal operating boundary} under constraints on target F1 and available review budget. Figure~\ref{fig:workload_vs_f1} plots, for every threshold pair \((\tau_l,\tau_u)\), the expected overall F1 (y–axis) against the fraction of items sent to human review (x–axis). Each gray dot corresponds to one policy \((\tau_l,\tau_u)\). The outlined curve is the \emph{Pareto frontier}: the set of optimal policies for which no other setting attains a higher F1 at the same (or lower) review load. Points strictly below/left of this curve are sub-optimal (i.e., worse F1 for equal or greater review).

Practically, the figure provides a budget-to-performance lookup: we can choose a review budget on the x-axis and read off the maximal achievable F1 on the frontier, along with the corresponding \((\tau_l,\tau_u)\). Under the Beta–mixture score regime shown here, for example, if the review budget is limited to 20\% of all decisions, the maximum attainable F1 score is approximately 0.93, given the optimal threshold values. As review increases, F1 improves but with diminishing returns. The curve marks efficient operating regions that balance accuracy gains against additional human effort. This frontier therefore identifies the feasible set of policies that are optimal under any monotone preference over accuracy and review cost. This particular analysis is conducted under the \emph{Beta–mixture} score regime.

Depending on the application, the F1 score may not always be the most appropriate performance metric. For example, in some domains, maintaining exceptionally high \emph{precision} is critical (such as in medical diagnosis). In other settings, such as threat detection, maximizing \emph{recall} is more important to ensure that few true positives are missed. Therefore, we also examine how the human review budget influences the two underlying components of the F1 score - \emph{precision} and \emph{recall}. 

\begin{figure}[t]
    \centering

    \subfloat[Workload vs precision\label{fig:workload_vs_precision}]{
        \includegraphics[width=0.47\linewidth]{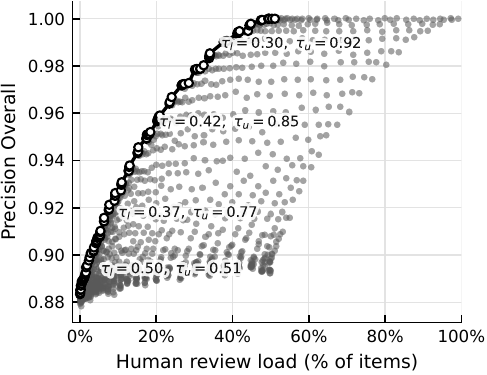}
    }
    \hfill
    \subfloat[Workload vs recall\label{fig:workload_vs_recall}]{
        \includegraphics[width=0.47\linewidth]{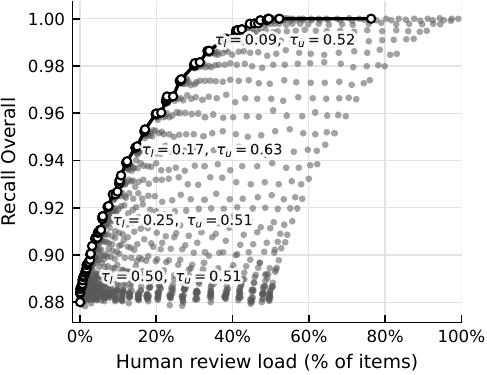}
    }
    
    \caption{Precision and recall trade-offs under varying human review budgets. 
    Each gray point represents an operating policy ($\tau_l$, $\tau_u$), while the 
    black-outlined curve marks the Pareto-optimal frontier. Annotated points 
    indicate representative threshold pairs.}
    \label{fig:workload_vs_precision_and_recall}
\end{figure}

Figure~\ref{fig:workload_vs_precision_and_recall} presents these complementary views: panel~\ref{fig:workload_vs_precision} plots the overall precision as a function of the review workload, while panel~\ref{fig:workload_vs_recall} shows the corresponding relationship for recall. Precision increases sharply with small amounts of human review and quickly saturates, while recall improves more gradually, benefiting from additional review of borderline cases near the decision thresholds. Analysis is conducted under the \emph{Beta–mixture} regime.

Together, these curves reveal the complementary dynamics between precision and recall as the review budget expands. Both curves exhibit diminishing returns beyond moderate review levels, identifying an efficient operating region in which marginal human effort yields the greatest overall benefit.

\begin{figure}[h]
    \centering
    \includegraphics[width=0.80\linewidth]{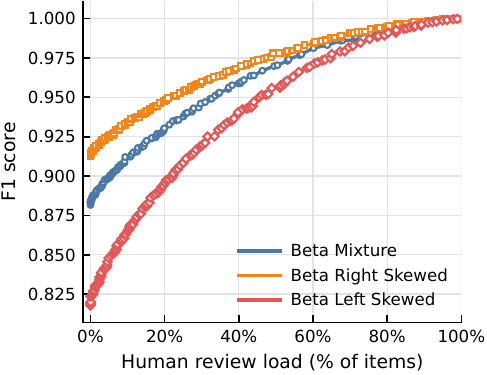}
    \caption{\textbf{Comparison of Pareto frontiers across score regimes.} Each curve shows the maximal achievable F1 score as a function of human review load under different simulated probability distributions. All curves exhibit diminishing returns beyond moderate review levels.}
    \label{fig:frontiers_f1}
\end{figure}

In Figure~\ref{fig:frontiers_f1}, we compare the Pareto frontiers for all three score regimes (\emph{Beta Mixture}, \emph{Beta Right Skewed}, and \emph{Beta Left Skewed}), showing how the best achievable F1 varies with review budget across settings. The comparison highlights how the underlying shape of the score distribution affects the possible trade-offs between automation and human oversight. In the right-skewed regime (representing a model where most predictions are confident) strong performance can be achieved with relatively little review. In contrast, the left-skewed regime, where predictions are generally less confident, requires substantially more human intervention to reach similar accuracy levels. Across all settings, the improvement curve exhibits diminishing returns: once a sufficient portion of uncertain cases is reviewed, further human effort yields only marginal F1 gains.

To further disentangle these effects, we next examine the evolution of \emph{precision} and \emph{recall} under varying review budgets, characterizing how each metric responds to changes in the lower and upper thresholds. Additional detailed results on precision and recall across all score regimes are provided in Figure~\ref{fig:frontiers_precision} and Figure~\ref{fig:frontiers_recall} in Appendix.

\subsection*{Empirical demonstration on a real entity-resolution system}
\begin{figure}[t]
\centering
\subfloat[Empirical score distribution\label{fig:empirical_distribution}]{
    \includegraphics[width=0.47\linewidth]{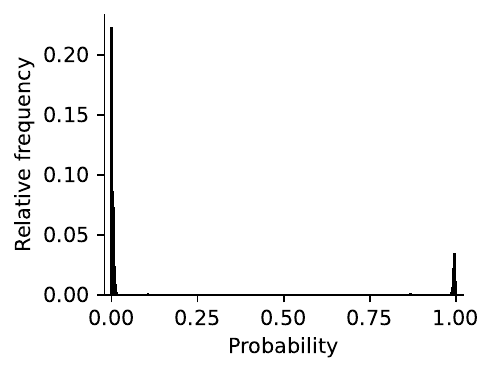}
}
\hfill
\subfloat[F1 vs. review load\label{fig:empirical_pareto}]{
    \includegraphics[width=0.47\linewidth]{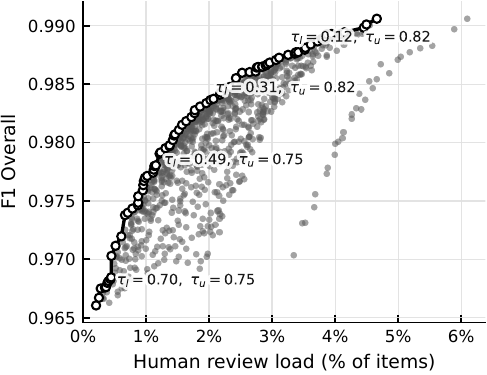}
}

\caption{\textbf{Empirical results on a real entity-resolution system.}
(a) Relative frequency of matching probabilities produced by our internal classifier over a large batch of candidate record pairs.
(b) Pareto frontier of expected F1 against human review load evaluated on the empirical distribution under the calibration assumption.}
\label{fig:empirical_results}
\end{figure}

To complement the simulation studies and demonstrate that the proposed framework operates as expected on real classifier outputs, we applied the same procedure to the matching probabilities produced by our internal entity-resolution system. Figure~\ref{fig:empirical_distribution} shows the empirical distribution of these scores over a large batch of candidate record pairs. The shape is consistent with the assumptions underlying our synthetic regimes. The bimodality is, however, strongly asymmetric and it most closely resembles the \emph{Beta Left-Skewed} case.

We do not possess ground-truth labels for this batch of pairs, which prevents a held-out evaluation in the conventional sense. We, however use the calibration assumption as defined in Section~\ref{sec:math_formulation}. It allows us to apply the same Monte Carlo procedure described in the previous subsections directly to the empirical scores.

Figure~\ref{fig:empirical_pareto} shows the resulting Pareto frontier of expected F1 against human review load. Two observations are worth highlighting. First, the qualitative behavior matches the predictions of the simulation studies. Second, the absolute operating regime is substantially more favorable than any of the synthetic cases studied above: the system attains F1 above $0.985$ while routing fewer than 3\% of pairs to human review, and approaches F1 near $0.99$ at review budgets below 5\%. We attribute this to the fact that the real classifier produces extremely confident predictions on the bulk of pairs. Taken together, these results corroborate the central claim of the simulation studies and demonstrate that the procedure is directly executable on real classifier outputs.

\section{Applications} \label{sec:applications}

The proposed framework applies broadly to any domain in which machine learning models produce probabilistic predictions that cannot be trusted blindly and thus require selective human oversight. In such settings, the model’s output probability represents its confidence, and the dual–threshold mechanism provides a principled way to determine when to automate a decision and when to defer it for human review. Below we outline several representative application areas.

In \textbf{entity resolution}, for instance, the system can automatically accept highly confident record matches, reject clearly distinct ones, and send uncertain pairs for manual review. In \textbf{fraud detection} and \textbf{financial auditing}, thresholding helps focus investigators on borderline cases rather than overwhelming them with false alarms. Similarly, in \textbf{medical image triage}, the approach allows confident positive or negative findings to be processed automatically, while ambiguous scans are referred to experts. Beyond these, the framework applies to \textbf{content moderation}, \textbf{misinformation detection}, \textbf{industrial inspection}, and \textbf{cybersecurity}.

\section{Discussion}

The framework operationalizes human--machine interaction as a constrained threshold-selection procedure derived directly from the Pareto frontier in Figure~\ref{fig:workload_vs_f1}. 

\textit{Implications for practitioners.} The simulations and the empirical demonstration illustrate patterns that practitioners can use to reason about their own deployments. First, the Pareto frontier exhibits clear diminishing returns across all regimes we examined: beyond an inflection point, the marginal gain per additional reviewed item becomes small. Second, the choice of objective function changes the optimal threshold structure: under calibrated probabilities, accuracy induces a symmetric review region centered near $p = 0.5$, whereas F1 induces an asymmetric region whose location depends on class prevalence. Based on the application (precision vs recall-sensitive) will inform the optimal thresholds.

\textit{Implications for researchers.} For the research community, the framework offers a reproducible framework for studying human--machine interaction policies in classification systems. Learned deferral approaches~\cite{mozannar2020consistent,cortes2023theory} can be evaluated as operating points relative to the post-hoc frontier produced by our procedure. A natural extension, inspired by curve-modeling approaches by Vatavu~\cite{vatavu2019}, is to fit a parametric model directly to the Pareto frontier of F1 versus review budget. The clear saturation pattern visible in Figure~\ref{fig:frontiers_f1} suggests that a small number of fitted parameters could summarize each score regime.

\subsection*{Limitations and future work}

Several limitations of the present framework merit acknowledgment. First, the formulation assumes calibrated probabilities. We expect the qualitative structure of the Pareto frontier to be robust to mild miscalibration, but practitioners should precede threshold selection with calibration diagnostics.

Second, we assume that human reviewers produce perfectly correct labels. This is a deliberate choice: the central goal of this work is to characterize how the threshold policy, the score distribution, and the review budget jointly determine system performance, and introducing reviewer error as an additional source of variability would obscure the structural properties of the framework that we set out to study. Extending to imperfect reviewers is straightforward but is left to future work.

Third, our empirical evaluation is restricted to a single entity-resolution system without adjudicated ground-truth labels. This is partly a consequence of the deployment context: production human-in-the-loop systems are typically proprietary, and labeled audit datasets are rarely released publicly. The empirical analysis should therefore be interpreted as a demonstration of the procedure's executability on real classifier outputs rather than as a held-out evaluation.

\subsection*{Reproducibility}
The code to reproduce all results is available at  
\href{https://github.com/gmuric/hil_sim}{\texttt{https://github.com/gmuric/hil\_sim}}.
\section*{Acknowledgment}
This work was sponsored by AFRL under Contract No.\ FA875025CB061.

\bibliographystyle{IEEEtran}
\bibliography{references}

\clearpage
\appendix


\begin{figure}[h]
    \centering
    \includegraphics[width=0.85\linewidth]{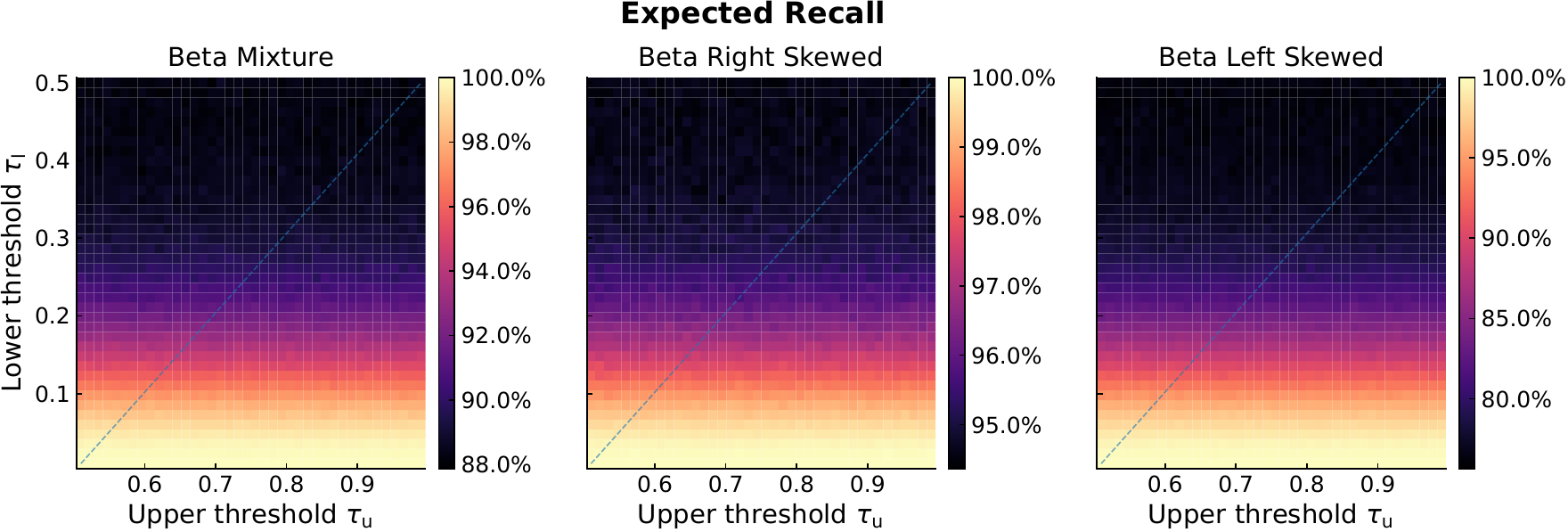}
    \caption{\textbf{Expected Recall.} Heatmaps showing the expected recall as a function of the lower ($\tau_l$) and upper ($\tau_u$) decision thresholds across three simulated score distributions. Recall generally increases as the lower threshold decreases, since more borderline positive cases are routed for human review.}
    \label{fig:RC_triptych}   
\end{figure}

\begin{figure}[h]
    \centering
    \includegraphics[width=0.85\linewidth]{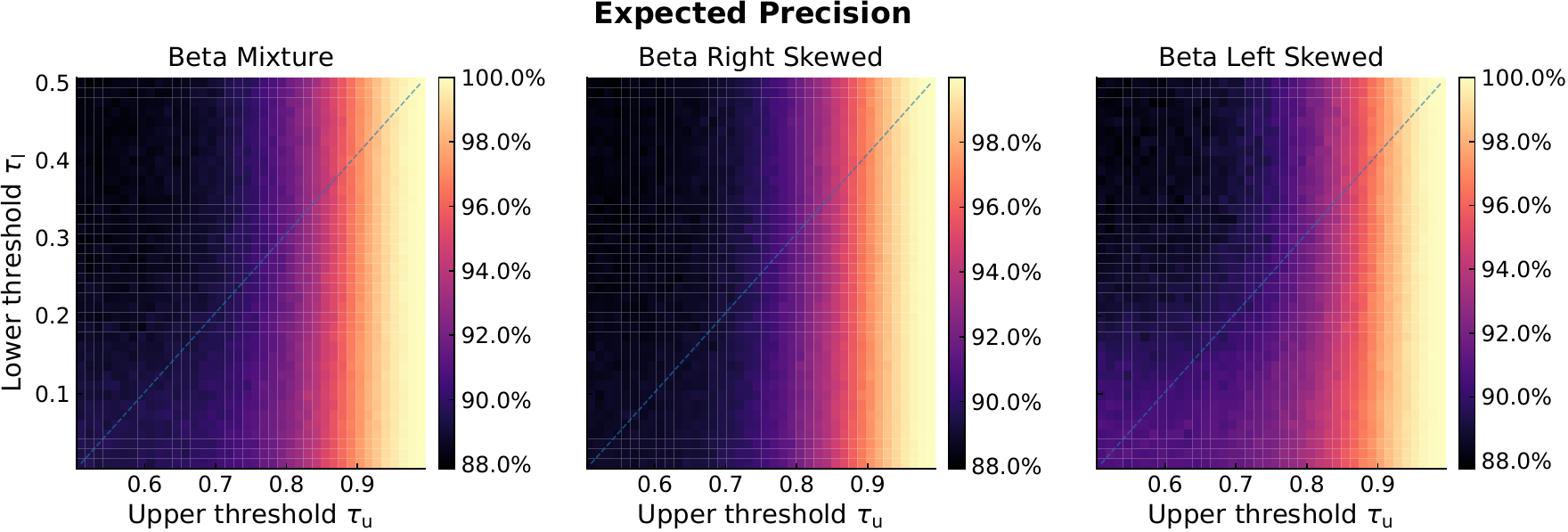}
    \caption{\textbf{Expected Precision.} Heatmaps showing the expected precision as a function of the lower ($\tau_l$) and upper ($\tau_u$) decision thresholds across three simulated score distributions. Precision increases primarily with higher upper thresholds, as the system becomes more conservative in automatically accepting positive classifications.}
    \label{fig:PR_triptych}
\end{figure}

\begin{figure}[h]
    \centering
    \includegraphics[width=0.85\linewidth]{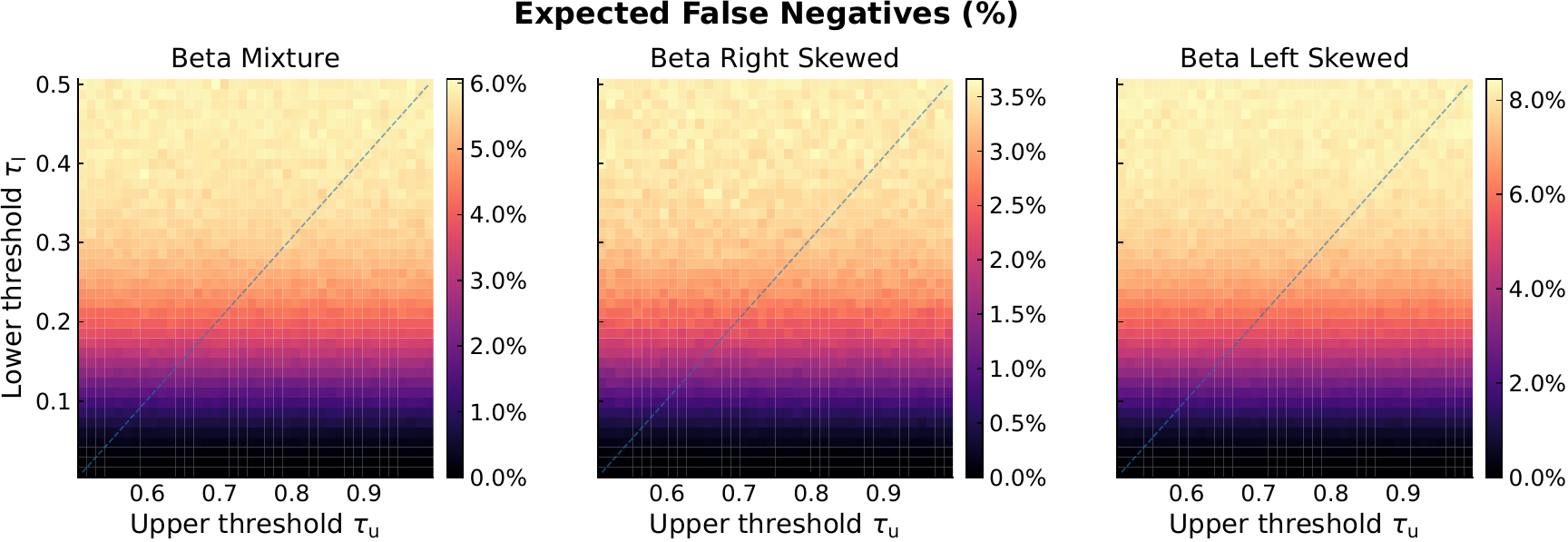}
    \caption{\textbf{Expected False Negatives (\%).} Heatmaps showing the expected false negative rate as a function of the lower ($\tau_l$) and upper ($\tau_u$) decision thresholds across three simulated score distributions. False negatives increase primarily with higher lower thresholds, as more true positives fall below the rejection boundary and are automatically classified negative.}
    \label{fig:FN_triptych}
\end{figure}

\begin{figure}[h]
    \centering
    \includegraphics[width=0.85\linewidth]{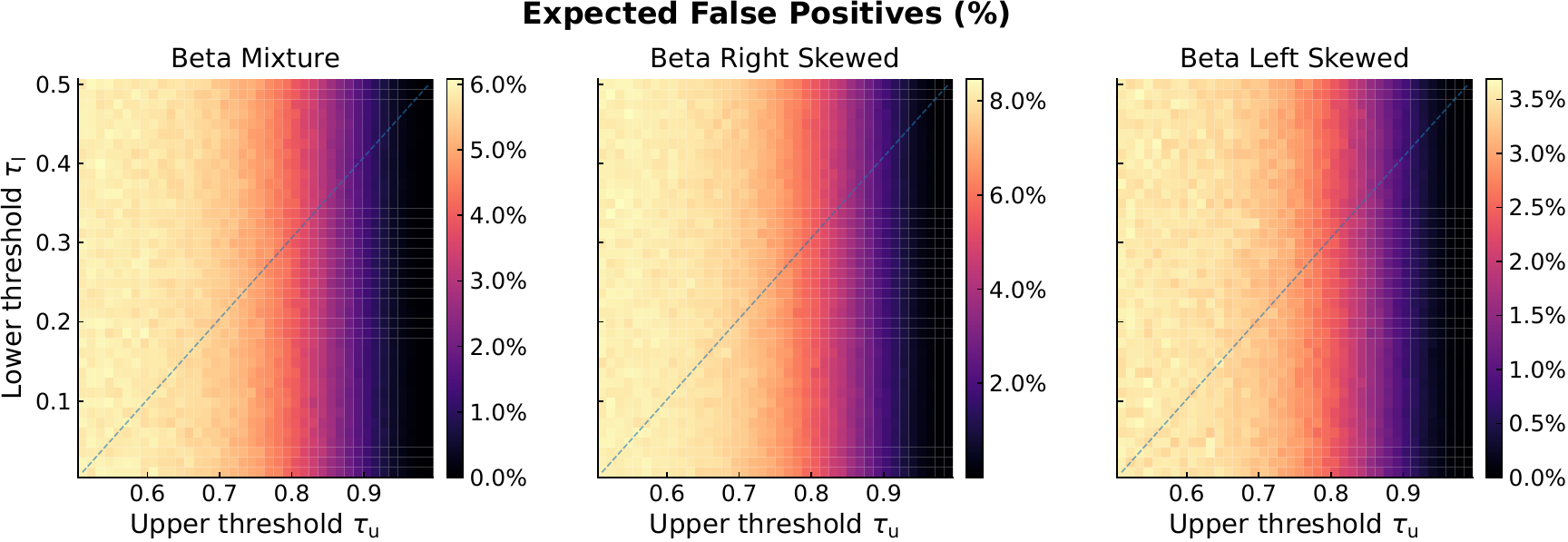}
    \caption{\textbf{Expected False Positives (\%).} Heatmaps showing the expected false positive rate as a function of the lower ($\tau_l$) and upper ($\tau_u$) decision thresholds across three simulated score distributions. False positives increase primarily when the upper threshold is set too low, allowing uncertain instances to be automatically classified as positive without review.}
    \label{fig:FP_triptych}
\end{figure}

\begin{figure}[h]
    \centering
    \includegraphics[width=0.85\linewidth]{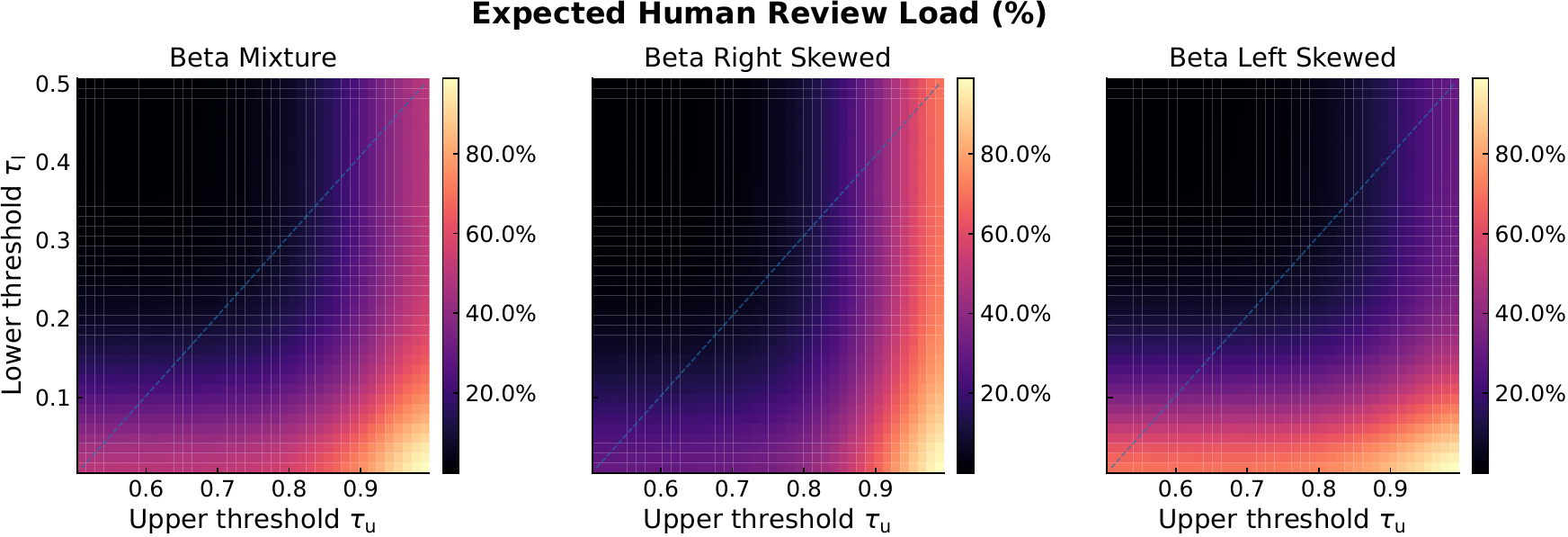}
    \caption{\textbf{Expected Human Review Load (\%).} Heatmaps showing the expected fraction of instances routed to human reviewers as a function of the lower ($\tau_l$) and upper ($\tau_u$) decision thresholds across three simulated score distributions. The human review load increases primarily when the gap between the two thresholds widens, expanding the uncertainty region where automated decisions are deferred.}
    \label{fig:human_load_triptych}
\end{figure}


    

\begin{figure}[h]
    \centering
    \includegraphics[width=0.85\linewidth]{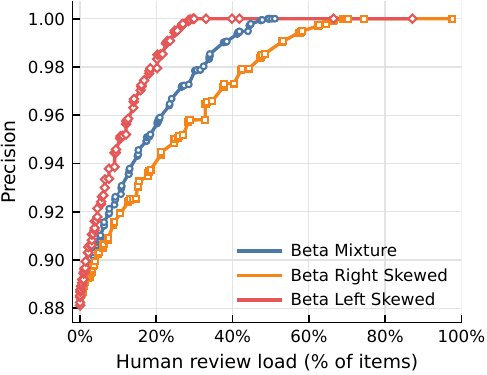}
    \caption{\textbf{Precision vs. Human Review Load.} Pareto frontiers showing the relationship between overall precision and the fraction of items sent for human review across three simulated score distributions. Precision improves rapidly with small amounts of human oversight and then plateaus, indicating diminishing returns beyond moderate review levels. 
    Systems with right-skewed score distributions (high model confidence) achieve high precision with minimal review, whereas left-skewed regimes require substantially more human intervention to reach comparable precision.}
    \label{fig:frontiers_precision}
\end{figure}

\begin{figure}[h]
    \centering
    \includegraphics[width=0.85\linewidth]{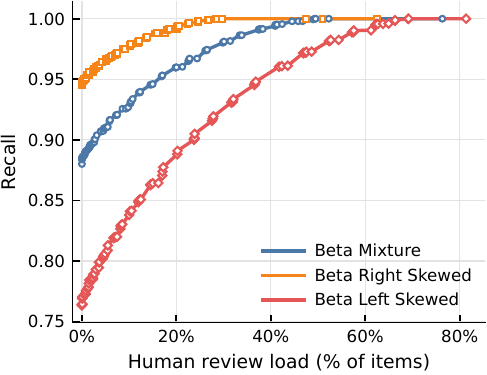}
    \caption{\textbf{Recall vs. Human Review Load.} Pareto frontiers showing the relationship between overall recall and the fraction of items sent for human review across three simulated score distributions. Recall increases steadily as the review workload expands, reflecting the recovery of true positives from the uncertain region near the decision thresholds. Systems with right-skewed score distributions achieve high recall with minimal review, while left-skewed regimes require substantially more human intervention to capture missed positives.}
    \label{fig:frontiers_recall}
\end{figure}

\end{document}